\documentclass[a4paper,11pt]{article}
\usepackage{jcappub} 
\usepackage{lineno}
\usepackage[utf8]{inputenc} 
\usepackage{newunicodechar} 
\newunicodechar{：}{:}
\DeclareUnicodeCharacter{00E9}{'e} 
\DeclareUnicodeCharacter{2009}{,}
\DeclareUnicodeCharacter{00A0}{~} 
\DeclareUnicodeCharacter{2013}{--} 
\DeclareUnicodeCharacter{2014}{---} 
\DeclareUnicodeCharacter{2026}{\ldots} 
\DeclareUnicodeCharacter{200B}{}
\usepackage{graphicx}
\usepackage{float}
\usepackage{amsmath}
\usepackage{bm}
\usepackage{amssymb}
\usepackage{nccmath}
\usepackage{placeins}
\usepackage[T1]{fontenc}
\usepackage{textcomp}
\usepackage{ulem}
\usepackage{tikz}

\usetikzlibrary{calc}

\title{\boldmath Implications~of~Cosmic~Birefringence \\ for Multi-Field ALP Dark Matter}

\author[a,c]{Jialiang Shao,}
\author[d,e]{Ippei Obata,}
\author[b,1]{Dongdong Zhang\note{Corresponding author}}

\affiliation[a]{Department of Astronomy, School of Physical Sciences, University of Science and Technology of China, Hefei 230026, China}
\affiliation[b]{School of Aerospace Information, Hefei Institute of Technology, Hefei 238706, China}
\affiliation[c]{CAS Key Laboratory for Research in Galaxies and Cosmology, School of Astronomy and Space Science, University of Science and Technology of China, Hefei 230026, China}
\affiliation[d]{Theory Center, Institute of Particle and Nuclear Studies (IPNS), High Energy Accelerator Research Organization (KEK), 1-1 Oho, Tsukuba, Ibaraki 305-0801, Japan}
\affiliation[e]{Kavli Institute for the Physics and Mathematics of the Universe (Kavli IPMU), WPI, UTIAS, The University of Tokyo, Kashiwa, Chiba 277-8568, Japan}

\emailAdd{jl210898@mail.ustc.edu.cn}
\emailAdd{iobata@post.kek.jp}
\emailAdd{don@hfit.edu.cn}

\abstract{Cosmic birefringence, characterized by the observed rotation of the polarization plane of the cosmic microwave background (CMB) radiation, serves as a critical probe for testing theories beyond the standard cosmological scenario. Axion-like particles (ALPs), particularly as a promising subcomponent of dark matter candidates, possess unique advantages in naturally explaining such phenomena. Previous studies on the implications of cosmic birefringence for these ultralight ALP fields have focused on single-field models with conventional potentials. These models face exclusion due to the washout effect—a suppression of the CMB polarization power spectrum induced by oscillatory dynamics of the scalar field within the mass range of $[10^{-25}, 10^{-22}]\text{ eV}$. To address this limitation, we develop a more general theoretical framework incorporating two ALP fields, providing analytical approximations and numerical calculations. Our findings reveal that while the general superposition of two ALP fields with distinct masses can't relax the constraints imposed by the washout effect and remains insufficient to fully explain the observed rotation angle at a 95\% confidence level. However, under specific conditions where the two fields are phase-aligned and their local amplitudes both reach the 95\% upper confidence limit, the two-field framework can theoretically account for the signal within the 1$\sigma$ observational uncertainty.}

\keywords{Cosmic Birefringence, Axion-like Particles, Ultralight Dark Matter}

\begin{document}

\tikz[remember picture, overlay] \node[anchor=north east, xshift=-1cm, yshift=-1cm] at (current page.north east) {\small Preprint: KEK-TH-2790, KEK-Cosmo-0405};

\maketitle
\flushbottom

\section{Introduction}\label{sec:intr}
{

The Cosmic Microwave Background (CMB) radiation, often regarded as the afterglow of the Big Bang approximately 380,000 years after its occurrence, provides a snapshot of the baryon–photon decoupling epoch on the last scattering surface. It encodes invaluable information about the physical conditions and subsequent evolution of the early universe.
In particular, studying the polarization of the CMB offers a sensitive probe of physical processes that have operated since the universe became transparent. 
Among these, cosmic birefringence \cite{Carroll:1989vb,Carroll:1991zs,Harari:1992ea,Nodland:1997cc} has recently gained considerable attention, especially given the growing need to test physics beyond the $\Lambda$CDM model.

Cosmic birefringence manifests as a rotation in the polarization direction of CMB photons during their journey from the last scattering surface to present-day observers. As a result, part of the $E$-mode polarization is converted into $B$-mode polarization, thereby mixing the $E$- and $B$-modes that would otherwise be statistically uncorrelated and generating a non-zero $EB$ cross-power spectrum \cite{Lue:1998mq,Feng:2006dp,PhysRevLett.103.051302,PhysRevD.89.063010,PhysRevD.89.103518,Lasky:2015lej,Zhao:2015mqa}. 
Such mixing serves as a sensitive diagnostic of potential parity-violating physics in the universe.
Current observations of the CMB $EB$ power spectrum have suggested a possible nonzero isotropic birefringence signal \cite{Minami_2020,PhysRevLett.128.091302,Eskilt_2022,Eskilt_20221,Cosmoglobe:2023pgf,Komatsu_2022}. 
For instance, joint analysis of $Planck$ and WMAP missions reported a polarization rotation angle $\beta = 0.34^\circ \pm 0.09^\circ$ with a $3.6\,\sigma$ statistical significance. Similarly, the Atacama Cosmology Telescope (ACT) Data Release (DR) 6 \cite{AtacamaCosmologyTelescope:2025blo,Diego-Palazuelos:2025dmh} obtained $\beta = 0.215^{\circ} \pm 0.074^{\circ}$, excluding $\beta = 0$ at $2.9\,\sigma$. Independent measurements, such as the POLARBEAR observations of the Crab Nebula \cite{Adachi_2024}, also support the existence of a time-varying birefringence signal. 

To explain the measured value of isotropic cosmic birefringence, a new particle which is not included in the Standard Model is necessary \cite{Nakai:2023zdr}.
Compelling theoretical interpretations of isotropic cosmic birefringence arise from the interaction between photons and axion-like particles (ALPs) as cosmological backgrounds such as dark energy \cite{Fujita:2020aqt,Choi:2021aze,Gasparotto:2022uqo}, early dark energy \cite{Fujita:2020ecn,Murai:2022zur,Eskilt:2023nxm}, topological defects \cite{Takahashi:2020tqv,Kitajima:2022jzz,Gonzalez:2022mcx,Ferreira:2023jbu,Jain:2021shf,Jain:2022jrp}, or coupled to dark matter \cite{Nakagawa:2021nme,Obata:2021nql}.
The rotation angle is given by a field displacement of the ALP field between the present and the position where CMB polarization was emitted. Owing to these properties, cosmic birefringence is sensitive to the ALP dynamics starting to oscillate in the history of the universe, which enables us to conduct tomographic tests of the cosmic birefringence angle at different redshifts caused by the ALP with mass $m_\phi \lesssim 10^{-28}$ eV \cite{Sherwin:2021vgb,Nakatsuka:2022epj,Namikawa:2025sft,Lee:2022udm,Namikawa:2023zux,Naokawa:2023upt,Yin:2024fez,Naokawa:2025shr,Naokawa:2024xhn}.

Among these parameter searches for ALPs in isotropic cosmic birefringence, its mass range over $10^{-27}$ eV has been less developed. For heavier mass range, the local oscillation of ALP field at the observer's location leads to time-dependent modulations in the observed polarization direction. 
Given that the oscillation period of ALP field scales inversely with its mass, for ALP mass $m_\phi \gtrsim 10^{-22}$ eV, the oscillation becomes the order of years to months comparable with the real time measurements and the birefringence angle is time-varying. Namely, a static birefringence signal requires much lighter ALPs, with $m_\phi \lesssim 10^{-22}\mathrm{eV}$. 
On the other hand, ALP with this mass range may experience a non-linear clustering following the gravitational instability associated with dark matter formation, and the field value can be amplified  compared with the cosmological average value for $m_{\phi} \lesssim 10^{-28}$ eV. 
While it is still unclear about a transition mass scale on which non-linear clustering becomes relevant, ALP with mass beyond $m_\phi \gtrsim 10^{-25}$ eV has been confirmed to experience it from the numerical simulation of mixed fuzzy and cold dark matter system \cite{Schwabe:2020eac,Lague:2023wes}.
Furthermore, inspired by ALPs from string theory \cite{Svrcek_2006,Conlon_2006,Arvanitaki:2009fg}, current cosmological observational data can be used to impose certain limitations on the mass range of axions \cite{PhysRevD.99.063517} . 
Therefore, our analysis focuses on the mass range between $10^{-25}\mathrm{eV}$ and $10^{-22}\mathrm{eV}$, where cosmic birefringence, structure formation and string theory constraints are relevant.

However, the presence of  ALPs with this mass range can introduce an additional effect in the CMB polarization, the so-called washout effect \cite{PhysRevD.79.063002,PhysRevD.100.015040,PhysRevD.111.043514}. 
The washout effect arises from rapid oscillations of the ALP field during the last scattering surface, which average out the rotation angle around its mean value $\overline{\beta}$ and reduce the net polarization amplitude. 
Recent studies \cite{zhang2024constrainingultralightalpdark} have shown that, under the assumption of a single ALP field, the coupling strength required to explain the observed birefringence angle is incompatible with the upper limits imposed by the washout effect for this mass range. 
This tension motivates the consideration of multi-field scenarios. 
String–inspired grand unified theories predict multiple ALP fields with independent masses and phases. 
Earlier works modeled their combined birefringence as a random walk, yielding a rotation angle scaling as the square root of the number of ALP fields $\sqrt{N}$ \cite{Gasparotto:2023psh,Gendler:2023kjt}.
On the other hand, for the washout effect, due to its different dependence on the ALP field, an appropriate distribution of fields may lead to a different scaling behavior and potentially relax the above tension.
Motivated by these considerations, this work investigates the theoretical constraints on cosmological birefringence and the washout effect within a two-field ALP framework.
Through analytic and numerical calculations, we show that the general superposition of two ALP fields can't alleviate the constraints that exclude the single-field explanation. Nevertheless, we find that if the two fields are phase-aligned and their amplitudes both reach the 95\% upper confidence level, the model can still theoretically explain the rotation angle within the 
1$\sigma$ observational error. This result opens up a broader parameter space and motivates further exploration of multi-field ALP frameworks.

This work is organized as follows.
In Section \ref{sec:cosmic birefringence}, we review the theoretical basis of birefringence and the washout effect in the single-field case.
In Section \ref{two fields}, we introduce the two-field ALP model, derive the corresponding constraints, and present our numerical results.
Finally, Section \ref{conclusion} summarizes our findings and discusses their broader implications for ALP dark matter and parity-violating cosmology.
Throughout this work,  we adopt natural units with $\hbar = c = 1$.

}

\section{Cosmic Birefringence and Axion-like Particles}\label{sec:cosmic birefringence}
{
In this section, we briefly review the derivation of the cosmic birefringence angle and washout effect from single ALP field dynamics.
ALP field can interact with photons via a Chern-Simons term in the action:
\begin{equation}
    S=\int d^4 x\sqrt{-g}(-\frac{1}{2}\nabla^{\mu}\phi\nabla_{\mu}\phi \, -V(\phi)-\frac{1}{4}F^{\mu\nu}F_{\mu\nu}-\frac{1}{4}g_{\phi\gamma}\phi F_{\mu\nu}\widetilde{F}^{\mu\nu}) .
\end{equation}
Here, $\phi$ denotes the ALP field, and $g_{\phi\gamma}$ represents the coupling strength between the ALP field and the electromagnetic field.
This interaction breaks the symmetry between the two circular polarization modes of photons, resulting in a rotation angle $\beta$ in the polarization direction of the photons. 
The magnitude of $\beta$ depends solely on the product of the coupling constant $g_{\phi\gamma}$ and the difference in the ALP field values between the last scattering surface and the observation point:
\begin{equation}
\begin{split}
    \beta(\hat{n}) &=\frac{1}{2}g_{\phi \gamma}(\phi(t_0,0)-\phi(t_{LSS},d_{LSS}\hat{n}))\\
    &= \frac{1}{2}g_{\phi \gamma}(\overline{\phi}_{obs}-\overline{\phi}_{LSS}+\delta\overline{\phi}_{obs}(\hat{n}))-\delta\overline{\phi}_{LSS}(\hat{n})\\
    &=\overline{\beta} \,+\delta\beta(\hat{n}) \,,
\end{split}
\end{equation}
where $t_0$ is the cosmic time at the observer, $\hat{n}$ is the unit direction vector, $t_{LSS}$ is the cosmic time at the last scattering surface, and $d_{LSS}$ is the comoving distance between the last scattering surface and the observer. 
$\overline{\beta}$  represents the isotropic component of the rotation angle, while $\delta\beta(\hat{n})$ denotes its anisotropic component which has been tightly constrained to near-zero levels by the $Planck$ satellite and ground-based observatories such as BICEP, SPT and ACT missions \cite{BICEP2:2017lpa,BICEPKeck:2022kci,Namikawa:2020ffr,SPT:2020cxx,Bortolami:2022whx,Zagatti:2024jxm,Namikawa:2024dgj}.
When the ALP mass is less than $10^{-22} \, \text{eV}$, the ALP de Broglie wavelength is much larger than the galactic scales. Within the Milky Way, such a field can be modeled as a wave packet composed of multiple modes with independent random phases, with the central value $\phi_0$ following a Rayleigh distribution \cite{Centers_2021}:
\begin{equation}
    P(\phi_{obs})=\frac{2\phi_{obs}}{\phi_0^2}\exp(-\frac{\phi_{obs}^2}{\phi_0^2}) .
\end{equation}
Assuming that the local energy density of ALP field $\rho_{0\phi}$ contributes a fraction $\kappa_\phi \equiv \rho_{0\phi}/\rho_{0dm}$ to the local dark matter energy density near the Milky Way $\rho_{0dm} \simeq 0.4 \, \text{GeV}/\text{cm}^3$ \cite{Iocco_2015,Sivertsson_2018}, it gives:
\begin{equation}
    \phi_0 \simeq 2.5\times 10^{12} \, \text{GeV} \times \kappa_{\phi}^{\frac{1}{2}}\times(\frac{m_\phi}{10^{-24}\,\text{eV}})^{-1} .
\end{equation}
Since the amplitude of $\phi_{obs}$ follows this Rayleigh distribution and phase follows a uniform distribution on $[0,2\pi]$, we can place an upper bound on $\phi_{obs}$ at the 95\% confidence level as $1.39\phi_0$ \cite{zhang2024constrainingultralightalpdark}. 
In reality, the emission of CMB photons spans a finite time interval $\Delta t_{\rm LSS}$, typically regarded as the full baryon-photon decoupling epoch, corresponding to redshifts between approximately 1200 and 1000, and a cosmic time duration of about $10^5$ years. Given that $t_{0} - t_{LSS} \gg \Delta t_{\rm LSS} \gg 2\pi/m_\phi$, the axion-like field undergoes many oscillations during the photon emission period, implying $\overline{\phi}_{LSS}\approx 0$. 
Consequently, the rotation angle $\beta$ is entirely determined by the local value of the ALP field at the observer’s location.
According to the latest measurements of ACT DR 6 \cite{Diego-Palazuelos:2025dmh}, the rotation angle is given by:
\begin{equation}
    \overline{\beta} =\frac{1}{2}g_{\phi \gamma}\phi_{obs} \simeq 0.215^{\circ}.
\end{equation}
So, under the ALP field explanation for birefringence, the coupling constant must satisfy at a 95\% confidence level :\textbf{
\begin{equation}
\begin{split}
  &g_{\phi\gamma} \gtrsim 2.16\times 10^{-15} \, \text{GeV}^{-1}\times \left(\frac{m_{\phi}}{10^{-24} \, \text{eV}}\right)\times \kappa_{\phi}^{-\frac{1}{2}} \times \left(\frac{\beta}{\overline{\beta}}\right) .
  \\&g_{\phi\gamma} \lesssim 6.64\times 10^{-15} \, \text{GeV}^{-1}\times \left(\frac{m_{\phi}}{10^{-24} \, \text{eV}}\right)\times \kappa_{\phi}^{-\frac{1}{2}} \times \left(\frac{\beta}{\overline{\beta}}\right) .\label{eq:2.6}
\end{split}
\end{equation}
}
However, it is essential to assess whether the associated washout effect will rule out the parameter region of cosmic birefringence or not.
In an expanding universe described by the FLRW metric $ds^2 = -dt^2 + a(t)^2d\textbf{x}^2$, ALP fields begin to oscillate once their mass exceeds the Hubble parameter. 
Under the WKB approximation, we have:
\begin{equation}
    \phi (t)  = \phi_0 \,a(t)^{-\frac{3}{2}}\cos(m_\phi t+\alpha(t))\, ,
\end{equation}
where $\alpha(t)$ is a phase factor which varies on the coherence time of ALP field: $\tau =2\pi/(m_\phi v^2)$.
Since CMB photons are emitted from a diffuse source randomly along the line of sight during the last scattering process,  we need to evaluate the time-averaged polarization rotation induced by the oscillating ALP field at the last scattering surface, $t = t_{\rm LSS}$.
The Stokes $Q$ and $U$ parameters are modified as \cite{PhysRevD.100.015040}:
\begin{align}
    (Q\pm i U)(\hat{n}) &= \epsilon I e^{\pm 2i(\psi+g_{\phi\gamma}\phi(t)/2)}\langle e^{\mp ig_{\phi\gamma}\phi(t_{\rm LSS})} \rangle \ , \notag \\
\langle e^{\mp ig_{\phi\gamma}\phi(t_{\rm LSS})} \rangle &\equiv \int dt'g(t')\exp\left[\mp i g_{\phi \gamma}\phi_\star\left(\frac{1+z(t^\prime)}{1+z_{\star}}\right)^{3/2} \cos(m_\phi t^\prime+\alpha(t'))\right] \ . \label{eq: wash1}
\end{align}
Here, $\epsilon$ represents the intrinsic polarization asymmetry of the source, $I$ is the total intensity of  polarization and $\psi$ is a fixed angle between polarization detectors and CMB polarization direction without ALP coupling.
The exponential term $e^{\pm ig_{\phi\gamma}\phi(t)}$ represents the local oscillation effect, arising from the oscillations of the axion field at the observer’s location\footnote{The rotation angle induced by this effect is tightly constrained by SPT-3G data; at the 95\% confidence level, its impact is limited to less than $0.071^{\circ}$ \cite{SPT-3G:2022ods} .
Since the local oscillation effect is based on observations made at the Earth's location, it is not subject to cosmic variance.
As such, it may provide even more stringent future constraints than the washout effect \cite{PhysRevD.100.015040}.}.
In the time integration part, $g(t)$ is the visibility function which has a peak at around $z=1000-1200$ with a time interval $\Delta t_{\rm LSS} \sim 10^5$ yrs, and $\phi_\star$ denotes the field value at the central redshift of recombination epoch $z = z_\star$.
As $|\partial_{t'} g(t')|\ll m_\phi$ , we can divide the integration into a large number of time domains $t' \in [t'_n - \delta t'/2, \ t'_n + \delta t'/2]$ with $\delta t'\equiv 2\pi/m_\phi$ so that in every interval $g(t_n') \simeq g_n$ and $z(t_n') \simeq z_n$ can be regarded as constant.
Regarding the phase factor, we can also assume a constant value $\alpha(t') \simeq \alpha'$ since the ALP coherence time scale is much longer than the time interval of last scattering surface for our interest mass range.
Then, we can analytically solve it as:
\begin{align}
\label{eq.washout}
\langle e^{\mp ig_{\phi\gamma}\phi(t_{\rm LSS})} \rangle &\simeq\sum_n g_n\int_{t_n^\prime - \delta t'/2}^{t_n^\prime + \delta t'/2}dt'\exp\biggl[ \mp i  g_{\phi \gamma} \phi_\star \left(\frac{1+z_n}{1+z_{\star}}\right)^{3/2} \cos(m_\phi t^\prime + \alpha') \biggr] \notag \\
&= \sum_n g_n\delta t' J_0\left( g_{\phi \gamma} \phi_\star\left(\frac{1+z_n}{1+z_{\star}}\right)^{3/2}\right) \notag \\
&= \int dt'g(t') J_0\left( g_{\phi \gamma} \phi_\star\left(\frac{1+z(t')}{1+z_{\star}}\right)^{3/2}\right) \ ,
\end{align}
where we have used a formula of Bessel function: $2\pi J_0(A) = \int^\pi_{-\pi}dxe^{i A\cos(x+\delta)}$.
Since $g(z)$ has a strong peak at $z = z_\star$, the weighted average of the ALP field amplitude in the integral is well approximated by its central value $\phi_\star$ in the integration.
Therefore, we finally obtain the analytical formula of observed Stokes parameters:
\begin{equation}
(Q\pm i U)(\hat{n}) \simeq e^{\pm ig_{\phi\gamma}\phi(t)}J_0\left( g_{\phi \gamma} \phi_\star\right)(Q\pm i U)_0(\hat{n}) \ ,
\end{equation}
where $(Q \pm i U)_0$ are Stokes parameters in the decoupling limit $g_{\phi\gamma} \rightarrow 0$.
According to the current simulation result based on $Planck$  \cite{PhysRevD.100.015040}, the washout effect imposes a constraint such that, at the 95\% confidence level, the reduction in polarization amplitude is limited to no more than 0.58\%:
\begin{equation}
    1-J_0 (g_{\phi \gamma} \phi_\star) \lesssim 5.8 \times 10^{-3} \, \ . \label{eq: wash_Planck}
\end{equation}
From the expansion $J_0(x) = 1 - x^2/4 + \mathcal{O}(x^4)$ for small $x$, we can see that the washout effect of ALP field exhibits a quadratic dependence on the polarization reduction, and this leads to the following upper bound on the coupling constant:
\begin{equation}
  g_{\phi\gamma} \lesssim 9.6\times 10^{-16} \, \text{GeV}^{-1} \times \left(\frac{m_{\phi}}{10^{-24}\,\text{eV}}\right) \times \left(f_{\phi} \times\frac{\Omega_{\rm DM} h^2}{0.11933}\right)^{-\frac{1}{2}} ,
\end{equation}
where $f_\phi \equiv \Omega_\phi/\Omega_{\rm DM}$ is a cosmological fraction of ALP dark matter abundance.

Regarding the formation of ALP dark matter through gravitational non-linear clustering, previous studies have shown that no significant change in the abundance fraction occurs for ALP masses $m_\phi \gtrsim 10^{-25}$ eV \cite{Schwabe:2020eac,Lague:2023wes}. We therefore assume $\kappa_\phi = f_\phi$ and compare the constraints in the parameter region of cosmic birefringence and the washout effect in Figure \ref{fig:placeholder}.
It can be seen that the upper limit on the coupling constant imposed by the washout effect is approximately two times lower than the lower bound required to account for the observed cosmological birefringence.
Therefore, single ALP model to explain cosmic birefringence has become less convincing in this mass range. 
\begin{figure}
    \centering
    \includegraphics[width=0.8\linewidth]{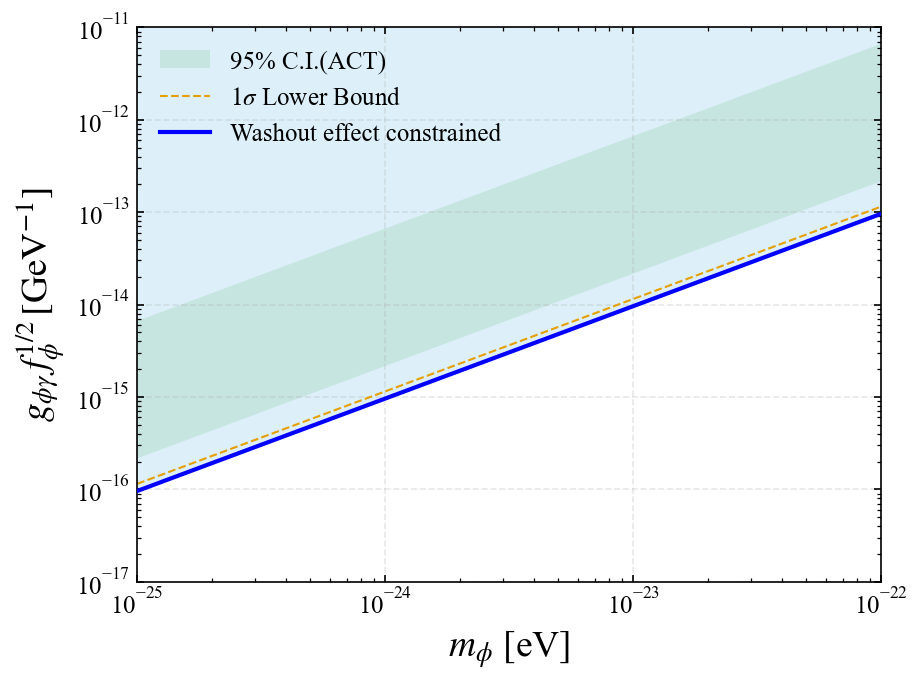}
    \caption{The constraints on the coupling constant from cosmic birefringence and Washout effect. The green shaded area represents the 95\% confidence interval for the coupling constant required to produce the observed cosmic birefringence (given $\beta = 0.215^\circ, \sigma = 0.074^\circ$), where the lower bound and upper bound of this region can be found in Equation \ref{eq:2.6}.\textbf{ }At the same time, the allowable parameter space should be below the blue solid line, so as not to be excluded by the Washout effect. The orange dashed line represents the 95\% confidence lower bound when the phase effect is not considered and the rotation angle is $\beta-\sigma$. It is easy to see that the washout effect's limitation on the coupling constant is ruled out by ACT observations within 1 $\sigma$ error range. In this plot, we assume that the local fraction of dark matter abundance is equal to the cosmological fraction of dark matter density: $\kappa_\phi = f_\phi$.} 
    \label{fig:placeholder}
\end{figure}

\section{Two ALP fields}\label{two fields}
{
In this work, we aim to perform a systematic calculation and analysis of a two-ALP field model. On one hand, we seek to resolve the issue that the coupling constant is too small to explain the observed cosmological birefringence due to the stringent washout effect constraint. On the other hand, this study also serves as a stepping stone for exploring more natural multi-field ($N$-field) theories.
In this section, we first derive the theoretical expressions in the two-field scenario and recalculate the conditions needed to account for the observed cosmic birefringence. We then derive approximate expressions for the washout effect, present our numerical results, and compare them with constraints from the single-field scenario.

\subsection{Theoretical Model}

We consider a model consisting of two independently evolving ALP fields that couple to photons with the same coupling constant $g_{\phi\gamma}$. The action is as follows:
\begin{equation}
\begin{split}
    S = \int d^4x \sqrt{-g} &\left( -\frac{1}{2} \nabla^\mu \phi^1 \nabla_\mu \phi^1 - \frac{1}{2} \nabla^\mu \phi^2 \nabla_\mu \phi^2 - \frac{1}{2} m_1^2 (\phi^1)^2 - \frac{1}{2} m_2^2 (\phi^2)^2 \right. \\
    &\left. - \frac{1}{4} F^{\mu\nu} F_{\mu\nu}  - \frac{1}{4} g_{\phi\gamma} (\phi^1 + \phi^2) F_{\mu\nu} \widetilde{F}^{\mu\nu} \right).
\end{split}
\end{equation}
Each field evolves independently as
\begin{equation}
    \phi^i(t) = \phi^i_{\text{ini}}\, a^{-3/2} \cos(m_i t + \alpha_i) \, ,
\end{equation}
parameterized by the initial amplitude $\phi^i_{\text{ini}}$, mass $m_i$ and phase factor $\alpha_i$ ($i=1,2$).
The amplitude of each field is characterized by its fractional contribution $\kappa_i$ to the total dark matter density:
\begin{equation}
    \phi_0^i \simeq 2.5 \times 10^{12} \, \text{GeV} \times \kappa_i^{1/2} \times \left( \frac{m_i}{10^{-24} \, \text{eV}} \right)^{-1}, \quad \kappa_1 + \kappa_2 = \kappa_{\phi} \,.
\end{equation}
To correctly calculate the cosmic birefringence effect caused by two independent axion fields, we need to sample over all the random variables. Considering that the amplitude of each field $\phi_{\text{obs}}^{(i)}$ follows a Rayleigh distribution within the coherence time, and the phases $\alpha_i'$ follow a uniform distribution in the range $[0, 2\pi]$, each field value $\phi_{\text{obs}}^{(i)}(t_0) = \phi_{\text{obs}}^{(i)} \cos \alpha_i'$ will follow a normal distribution. The total field value, which is equal to the sum of the two independent field values, also follows a normal distribution. Since the sign of the total field value must be the same as the cosmic birefringence angle, it follows a half-normal distribution. Consequently, the coupling constant $g_{\phi\gamma}$ is inversely proportional to the total field value, thus following an inverse half-normal distribution. This allows us to establish its upper and lower bounds at a 95\% confidence level:
\begin{equation}
\begin{split}
   &g_{\phi\gamma} \gtrsim 2.16 \times 10^{-15} \, \text{GeV}^{-1} \times\left(\frac{m_{1}m_2}{\sqrt{\kappa_1 m_2^2+\kappa_2m_1^2} }\right)\times\frac{1}{10^{-24}~\mathrm{eV}} \times \left( \frac{\beta}{0.215^\circ} \right),
\\&g_{\phi\gamma} \lesssim 6.64 \times 10^{-14} \, \text{GeV}^{-1} \times\left(\frac{m_{1}m_2}{\sqrt{\kappa_1 m_2^2+\kappa_2m_1^2} }\right)\times\frac{1}{10^{-24}~\mathrm{eV}} \times \left( \frac{\beta}{0.215^\circ} \right).
\label{eq:3.4}
\end{split}
\end{equation}
However, under specific conditions where the two fields are phase-aligned and their local amplitudes both reach the 95\% upper confidence limit, we can obtain the minimum values of the coupling constant allowed by the two-axion theory:
\begin{equation}
\begin{split}
    g_{\phi\gamma,min} &= \frac{2\beta}{1.73(\phi_0^1+\phi_0^2)}  \\
    &=1.72\times 10^{-15} \text{GeV}^{-1}\times \left(\kappa_1^{\frac{1}{2}}\,(\frac{m_1}{10^{-24}\text{eV}})^{-1}+\kappa_2^{\frac{1}{2}}\,(\frac{m_2}{10^{-24}\text{eV}})^{-1}\right)^{-1}\times \left( \frac{\beta}{0.215^\circ} \right).
\end{split}
\end{equation}
One simple realization of  phase alignment could be achieved by considering the late-time interaction with the dynamical dark energy field, which may induce such effective potential $V_{int}\propto\rho_{de}\cos{(\phi_1-\phi_2)} $ and force the two axion fields to be aligned. A concrete\textit{ EFT } construction is, however, nontrivial and beyond the scope of this paper. We leave it for future studies. And in the situation where the phases are only partially aligned, we would expect this equation to gradually return to the case of random
phases. To compute the washout effect in the two field model, the previous formula \eqref{eq.washout} is not strictly applicable, as it requires tracking the evolution of the multiple oscillation phases over time.
We can write the expression as:
\begin{align}
    &(Q\pm i U)(\hat{n}) = \epsilon I e^{\pm 2i(\psi+g_{\phi\gamma}(\phi^1(t)+\phi^2(t))/2)}\langle e^{\mp ig_{\phi\gamma}(\phi^1(t_{\rm LSS})+\phi^2(t_{\rm LSS}))} \rangle \ , \notag \\
&\langle e^{\mp ig_{\phi\gamma}(\phi^1(t_{\rm LSS})+\phi^2(t_{\rm LSS}))} \rangle \notag \\
&\equiv \int dt'g(t') \exp \left[ \mp i 
g_{\phi\gamma} \left( \frac{1 + z(t')}{1 + z_\star} \right)^{3/2} \left( \phi_\star^1 \cos(m_1 t' + \alpha_1') + \phi_\star^2 \cos(m_2 t' + \alpha_2') \right) \right] \notag \\
&=\int dt'g(t')\sum_{n=-\infty}^\infty\sum_{m=-\infty}^\infty(\mp i)^{n+m} \notag \\
&\times J_n\left[g_{\phi\gamma}\phi^1_\star\left( \frac{1 + z(t')}{1 + z_\star} \right)^{3/2}\right]J_m\left[g_{\phi\gamma}\phi^2_\star\left( \frac{1 + z(t')}{1 + z_\star} \right)^{3/2}\right]e^{i\left((n\,m_1+m \,m_2)t'+n \alpha_1'+m\alpha_2'\right)} \ , \label{eq: wash2}
\end{align}
where we have used Jacobi-Anger identity $e^{\pm iA\cos\theta}=\sum_{n=-\infty}^\infty(\pm i)^nJ_n(A)e^{in\theta}$.
Since $g_{\phi\gamma}\phi^i \ll 1$, we can expand the Bessel function and keep the lowest order of $g_{\phi\gamma}\phi^i$.
Then, due to a rapid oscillation of ALP field in the phase sector with $nm_1+mm_2 \neq 0$, only the contribution from $J_0$ remains in the integration:

\begin{align}
\langle e^{\mp ig_{\phi\gamma}(\phi^1(t_{\rm LSS})+\phi^2(t_{\rm LSS}))} \rangle &\simeq \int dt'g(t') J_0\left[g_{\phi\gamma}\phi^1_\star\left( \frac{1 + z(t')}{1 + z_\star} \right)^{3/2}\right]J_0\left[g_{\phi\gamma}\phi^2_\star\left( \frac{1 + z(t')}{1 + z_\star} \right)^{3/2}\right] \notag \\
&\simeq J_0\left(g_{\phi\gamma}\phi^1_\star\right)J_0\left(g_{\phi\gamma}\phi^2_\star\right) \ . \label{eq: wash3}
\end{align}
Finally, the observed Stokes parameters are approximately given by
\begin{equation}
    (Q\pm i U)(\hat{n}) \simeq e^{\pm ig_{\phi\gamma}(\phi^1(t)+\phi^2(t))}J_0\left(g_{\phi\gamma}\phi^1_\star\right)J_0\left(g_{\phi\gamma}\phi^2_\star\right)(Q\pm i U)_0(\hat{n}) \label{eq: wash4} \ .
\end{equation}
Following \eqref{eq: wash_Planck}, the constraint on the washout effect from two ALP fields reads
\begin{equation}
1-J_0\left(g_{\phi\gamma}\phi^1_\star\right)J_0\left(g_{\phi\gamma}\phi^2_\star\right) \simeq \frac{g_{\phi\gamma}^2}{2}\left(\frac{f_1}{m_1^2}+\frac{f_2}{m_2^2}\right)\, \rho_{\star \mathrm{DM}} \lesssim 5.8\times10^{-3} \ , \quad f_1 + f_2 = f_\phi \ .
\end{equation}
This leads to the following upper bound on the coupling constant:
\begin{equation}
\label{eq:washout}
       g_{\phi\gamma} \lesssim  9.6 \times 10^{-16}~\mathrm{GeV}^{-1}\times \left(\frac{m_{1}m_2}{\sqrt{f_1 m_2^2+f_2m_1^2} }\right)\times\frac{1}{10^{-24}~\mathrm{eV}}\times \left( \frac{\Omega_{\rm DM} h^2}{0.11933}\right)^{-\frac{1}{2}} \ .
\end{equation}
Regarding the contribution from Bessel functions at the quadratic order of $g_{\gamma\phi}\phi$ in \eqref{eq: wash2}, we notice that another term $J_1(g_{\gamma\phi}\phi)J_{-1}(g_{\gamma\phi}\phi) = \mathcal{O}(g_{\gamma\phi}^2\phi^2)$ could remain with $(m_1-m_2)t' \simeq 0$.
However, when the condition $|m_1 - m_2|\gg 2\pi/\Delta t_{\rm LSS}$ is satisfied, the frequency separation between the two fields is much larger than the inverse of the duration of recombination. Since  $2\pi/\Delta t_{\rm LSS}$ is approximately on the order of $10^{-27}$ eV, the condition $|m_1 - m_2|\gg 2\pi/\Delta t_{\rm LSS}$ can be satisfied in most of the parameter space.
As a result, the coherence between the two modes decays rapidly, allowing this term to be neglected after the integration.

\subsection{Numerical Calculation}

To obtain more accurate and intuitive results, we perform direct numerical integration of the full expression. According to the most stringent constraints from CMB and large-scale structure observations, ALP can constitute the entirety of dark matter only if their mass exceeds $10^{-24}~\mathrm{eV}$.
Otherwise, it can contribute only a fraction of the total dark matter density~\cite{Hlozek_2015,Hlo_ek_2018,Lagu__2022,Rogers_2023}. 
More recently, measurements of the UV luminosity functions of high-redshift galaxies have constrained the fractional contribution of ultra-light ALPs with masses in the range $10^{-25}~\mathrm{eV}$–$10^{-18}~\mathrm{eV}$ to about 1\% of the total dark matter~\cite{winch2025highredshiftsmallscaletestsultralight}.
In this work, we assume that the ALP fields constitute 1\% of the total dark matter energy density $f_{\phi} = 0.01$.
If we can assume $\kappa_i = f_i$ and hence $\kappa_\phi = f_\phi$, since both cosmic birefringence and the washout effect depend on $f_{\phi}$  in the same manner, varying $f_{\phi}$ does not qualitatively alter our results.
For the visibility function $g(t(z))$, we adopt a Gaussian profile centered at redshift $z=1089$. 
Then, choosing the span of redshift $z_{\text{span}}=60$ ensures that $g(t)$ evolves slowly over the relevant timescale.

In Figure~\ref{fig:two_images_equal}, we fix the mass of one ALP field $\phi^1$ and show the parameter constraints when the two fields contribute equally to the dark matter density. We find that the general superposition of two ALP fields can't relax the constraint that excludes the single-field explanation, so we focus on the specific superposition in which phases are aligned and amplitudes both reach the 95\% upper confidence level in the following discussion.\textbf{ }In this case, the two-field model admits a more relaxed parameter space relative to the single-field case: the ratio of tension is reduced roughly by a factor of $\sqrt{2}$  , remaining consistent with observations within the 1$\sigma$ uncertainty. 
\begin{figure}[t]
\centering

\includegraphics[width=1.0\linewidth]{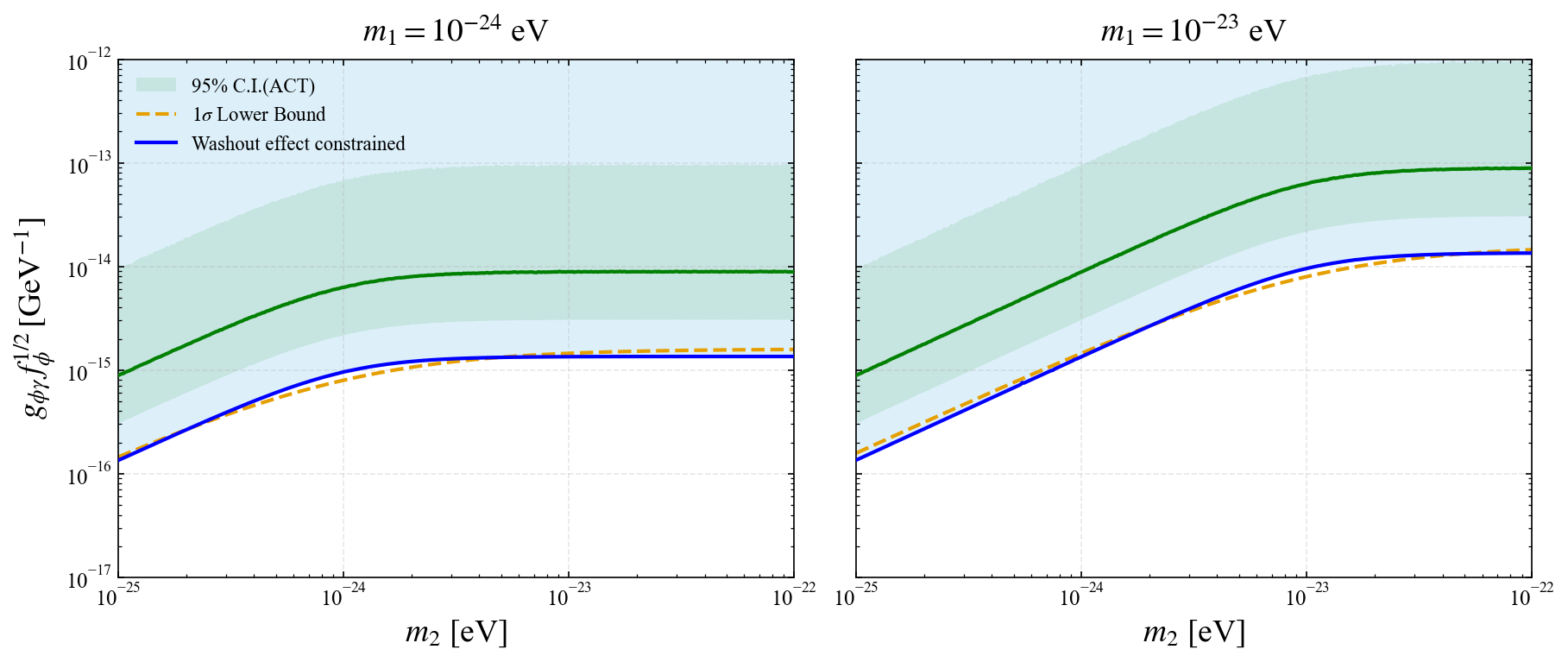}

\caption{  Constraints on the coupling constant from cosmic birefringence and the washout effect. Here $f_1 = f_2 = 0.5f_\phi$ with both ALP fields contributing equally to the total dark matter density. The panels correspond to $m_1 = 10^{-24}\text{ eV}$ and $m_1 = 10^{-23}\text{ eV}$ from left to right, respectively. The horizontal axis represents $m_2$. The green shaded area represents the 95\% confidence interval for the coupling constant required to produce the observed cosmic birefringence (given $\beta = 0.215^\circ, \sigma = 0.074^\circ$). The solid green line represents the coupling constant corresponding to the points of maximum probability within this interval. The orange line represents the minimum values required by the two-axion fields theory to produce corresponding rotation angles ($\beta - \sigma$). The solid blue line denotes the limit imposed by the washout effect, and the light blue area above it is excluded.}
\label{fig:two_images_equal}
\end{figure}

We further consider scenarios in which the two fields contribute unequally to the dark matter density in Figure \ref{fig:unequal_case}.
When the fraction associated with the $m_1$ field decreases, a viable parameter region remains within the 1$\sigma$ uncertainty. However, for fixed $m_1$, the allowed region shifts toward larger masses. In the extreme limit where the $m_1$ field contributes much less than the $m_2$ field, the two-field model is excluded by the washout effect within 1$\sigma$ across the entire mass range. In this case, the behavior effectively reduces to that of the single-field scenario.
\begin{figure}[t]
\centering

\includegraphics[width=1.0\linewidth]{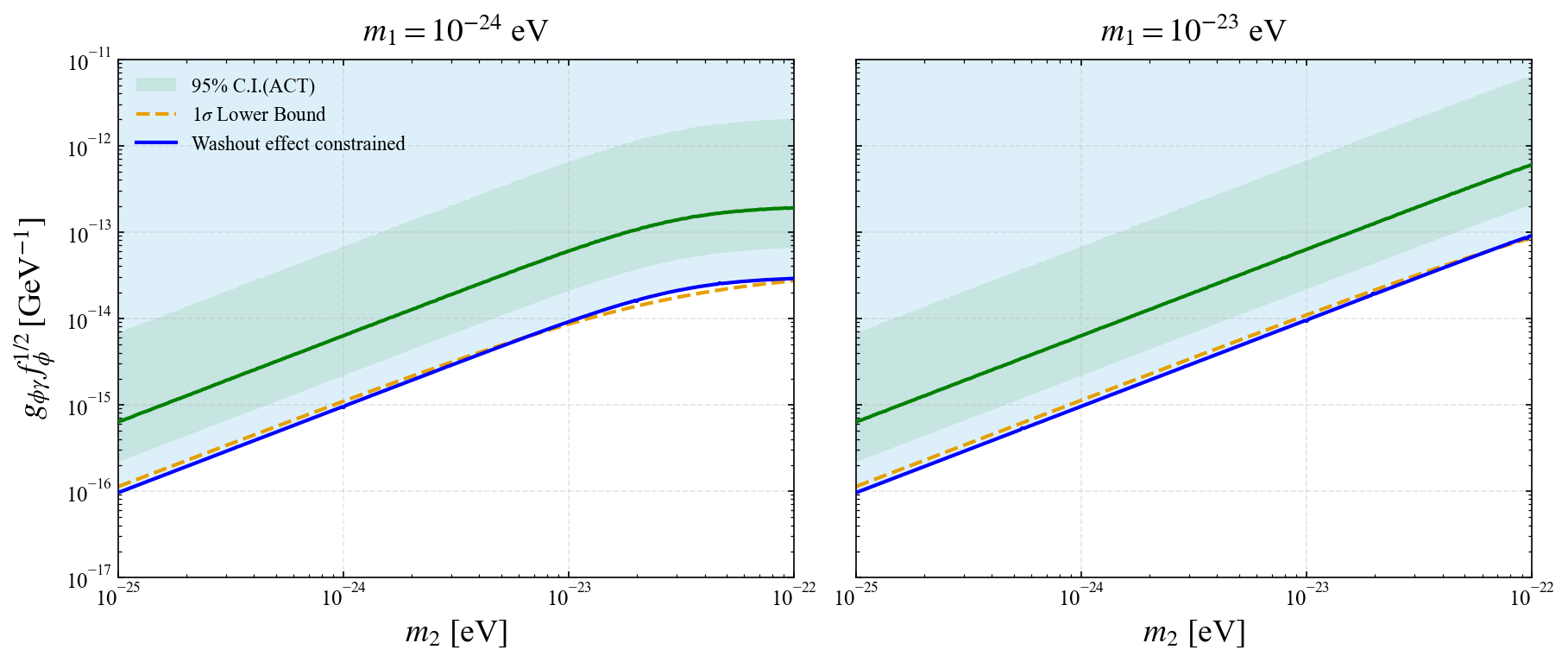}

\caption{Same as Figure 2  but with different $f_1$ and $f_2$. $f_1=0.001 f_{\phi}$, $f_2=0.999 f_{\phi}$.}
\label{fig:unequal_case}
\end{figure}

To ensure the robustness of our conclusions, we also incorporate the constraints from WMAP and $Planck$ measurements of CMB polarization~\cite{Eskilt:2022cff}. Specifically, we use the combined $Planck$ HFI + LFI + WMAP data, including the filamentary dust $EB$ model and allowing nonzero calibration angle offsets. In this case, the observed rotation angle is $\beta = 0.34^\circ \pm 0.09^\circ$.
In Figure \ref{fig:planck_constraints}, while the two-field model does not fully account for the observed static cosmic birefringence within the $2 \sigma$ constraints from the data, the tension gets more relaxed than the single-field model. 
However, our determination of the coupling constant depends on the local dark matter density. 
Taking into account the uncertainties in measurements of the local dark matter density, arising from data limitations, the assumption of a flat rotation curve, or possible disequilibrium effects \cite{Sivertsson:2017rkp}, we find that the two-field theory remains compatible with the Planck joint data at the $2 \sigma$ level.  
\begin{figure}[t]
\centering
\begin{minipage}{0.48\textwidth}
\centering
\includegraphics[width=\textwidth]{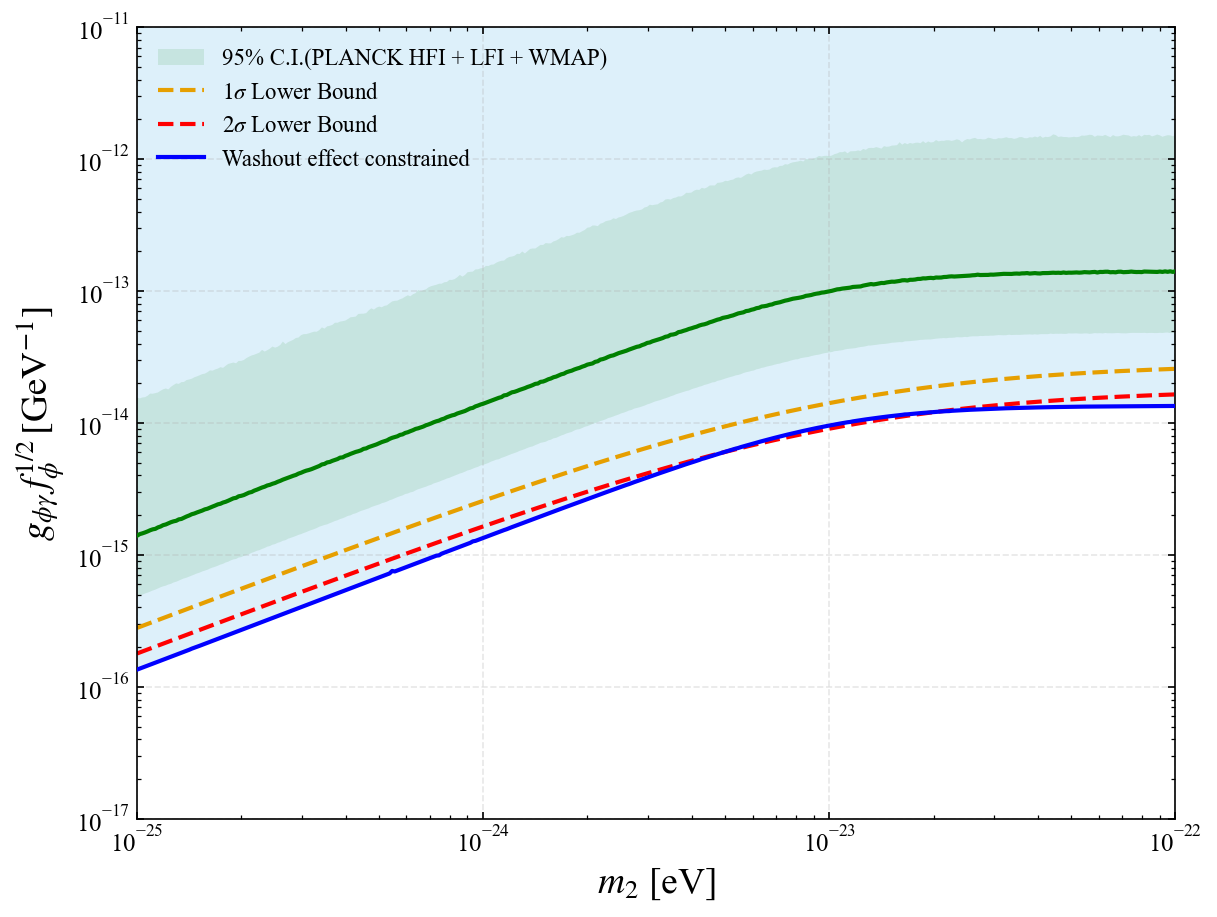}
\end{minipage}
\hfill
\begin{minipage}{0.48\textwidth}
\centering
\includegraphics[width=\textwidth]{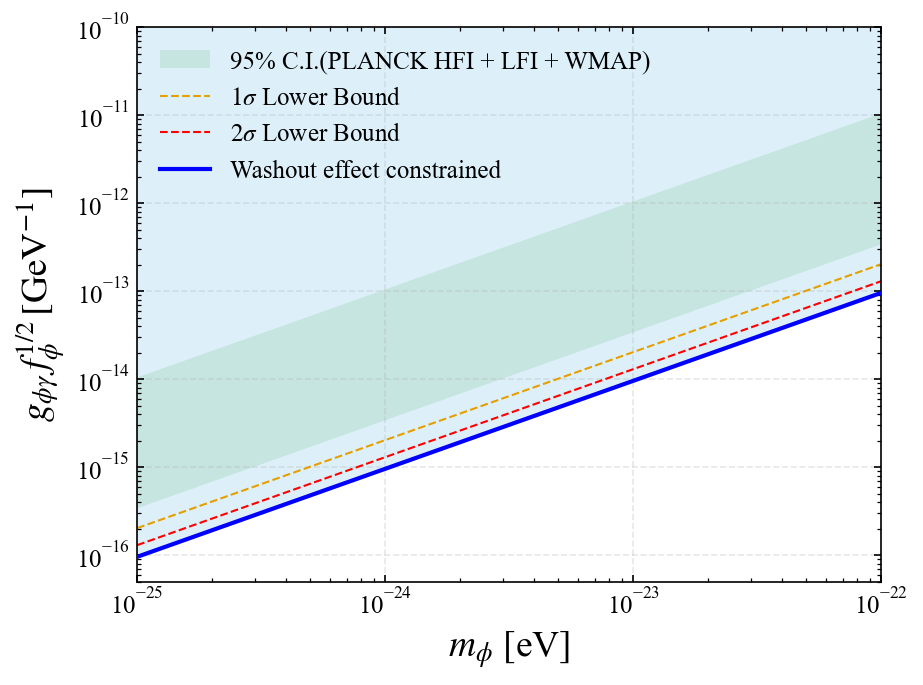}
\end{minipage}
\caption{Constraints from the combined Planck HFI + LFI + WMAP data. The left panel shows the two-field model with $f_1=f_2=0.5 f_{\phi}$ and $m_1=10^{-23}\,\mathrm{eV}$ , while the right panel displays the corresponding single-field case.  The orange and red dashed lines are  the minimum values required by the two-axion fields theory to produce corresponding rotation angles ($\beta - \sigma$ and $\beta - 2\sigma$, respectively)}
\label{fig:planck_constraints}
\end{figure}

\section{Discussion and Conclusion}\label{conclusion}
{
We first briefly review the phenomenon of cosmic birefringence and proposal that ALP could potentially account for the observed effect for its masses $m_\phi \gtrsim 10^{-25}$ eV where ALP could behave as fuzzy dark matter.
However, the introduction of such ALPs inevitably induces additional physical effects on the CMB polarization, namely the washout effect.
We recall that, under these new constraints, the coupling strength permitted by single-field ALP models is insufficient to reproduce the observed static cosmic birefringence. Consequently, such models are ruled out by current observational data. This naturally raises the question of whether introducing two or more ALP fields could alleviate or even resolve the difficulties encountered in the single-field scenario.

We then derive a semi-analytical expression for the upper bound on the interaction coupling constant in the two-field framework. Our rigorous statistical analysis indicates that at a 95\% confidence level, the two-field model is still insufficient to fully explain the central value of the cosmic birefringence signal from ACT DR6. Nevertheless,  assuming the two fields are phase-aligned and their local amplitudes both reside at the 95\% upper confidence limit, the theory can theoretically account for the observation within the 1$\sigma$ error margin. In this specific scenario, the constraint primarily depends on two key factors. The first factor is the mass hierarchy between the two fields. When masses are comparable, the constraint from the washout effect is relaxed, alleviating the tension compared to a single-field case. 
The second factor is the fraction hierarchy. Once the fraction of one field falls below 0.1\%, two-field model will obtain similar constraint results to the single-field model.
We further present numerical results and compare them with the single-field constraint curves to verify the validity of our theoretical estimates. Our findings indicate that the two-field model can indeed relax the washout effect constraints and explain the cosmic birefringence signal from ACT DR6 and potentially from $Planck$/WMAP data.

More dedicated search for the probability distribution of parameter space leaves in our future work.
Our calculations also rely on measurements of the dark matter density at the observer's location. If ALPs could constitute a higher proportion of the dark matter in the solar system \cite{Cuadrat-Grzybowski:2024uph,Budker:2023sex}, then the number of ALP fields needed to explain cosmic birefringence would decrease accordingly.

}

\section*{Acknowledgments}
We are grateful for Yi-Fu Cai, Elisa Ferreira, Kai Murai and Fumihiro Naokawa for insightful discussions. 
This work is supported in part by the Youth Project of Scientific Research Program of Anhui Provincial Department of Education (2025AHGXZK40234), by the Hefei Institute of Technology Talent Research Fund (2025KY41), by the National Key R\&D Program of China (2021YFC2203100, 2024YFC2207500), by the National Natural Science Foundation of China (12433002, 12261131497, 92476203), by CAS young interdisciplinary innovation team (JCTD-2022-20), by 111 Project (B23042), by USTC Fellowship for International Cooperation, by USTC Research Funds of the Double First-Class Initiative and by JSPS KAKENHI Grant No. 19K14702 (IO).

\bibliographystyle{apsrev4-1}
\bibliography{references}
\end{document}